\documentclass{aa}
\usepackage{graphicx}
\usepackage{xcolor}
\usepackage{booktabs}
\usepackage{supertabular}
\usepackage[version=4]{mhchem}
\usepackage{ulem}
\usepackage{longtable}
\usepackage{caption}
\usepackage{sidecap}
\usepackage{float}
\usepackage{placeins}
\usepackage{rotating}
\usepackage{txfonts}
\usepackage{hyperref}
\hypersetup{
    colorlinks=true,
    linkcolor= blue,
    filecolor=magenta,      
    urlcolor=blue,
    citecolor= blue,
}%
\begin{document}

   \title{Hunt for complex cyanides in protostellar ices with JWST}

   \subtitle{Tentative detection of CH$_3$CN and C$_2$H$_5$CN}

   \author{P. Nazari
          \inst{1}
          \and
          W. R. M. Rocha\inst{2}
          \and
           A. E. Rubinstein\inst{3}
           \and
           K. Slavicinska\inst{1}
           \and
           M. G. Rachid\inst{2, 4}
           \and
           E. F. van Dishoeck\inst{1,5}
           \and
           S. T. Megeath\inst{6}
           \and
           R. Gutermuth\inst{7}
           \and
           H. Tyagi\inst{8}
           \and
           N. Brunken\inst{1}
           \and
           M. Narang\inst{8,9}
           \and
           P. Manoj\inst{8}
           \and
            D. M. Watson\inst{3}
           \and
           N. J. Evans II\inst{10}
           \and
           S. Federman\inst{6}
           \and
           J. Muzerolle Page\inst{11}
           \and
           G. Anglada\inst{12}
           \and
           H. Beuther\inst{13}
           \and
           P. Klaassen\inst{14}
           \and
           L. W. Looney\inst{15}
           \and
           M. Osorio\inst{12}
           \and
           T. Stanke\inst{5}
           \and
           Y.-L. Yang\inst{16}
           %\and
           %Others
           % P. Atnagulov\inst{14}
           % \and
           % T. L. Bourke\inst{15}
           % \and
           % A. Caratti o Garatti\inst{16}
           % \and
           % W. J. Fischer\inst{17}
           % \and
           % E. Furlan\inst{18}
           % \and
           % J. Green\inst{17}
           % \and
           % N. Habel\inst{19}
           % \and
           % L. Hartmann\inst{20}
           % \and
           % N. Karnath\inst{21, 22}
           % \and
           % H. Linz\inst{11, 23}
           % \and
           % L. W. Looney\inst{24, 25}
           % \and
           % J. Muzerolle\inst{17}
           %  \and
           %  R. Pokhrel\inst{14}
           %  \and
           % R. Rahatgaonkar\inst{26}
           % \and
           % P. Sheehan\inst{25}
           % \and
           % T. Stanke\inst{5}
           % \and
           % A. M.\ Stutz\inst{27}
           % \and
           % J. J. Tobin\inst{25}
           % \and
           % {\L} Tychoniec\inst{28}
           % \and
           % S. Wolk\inst{22}
          }

   \institute{Leiden Observatory, Leiden University, P.O. Box 9513, 2300 RA Leiden, Netherlands\\ 
        \email{nazari@strw.leidenuniv.nl}
         \and
         Laboratory for Astrophysics, Leiden Observatory, Universiteit Leiden, Leiden, Zuid-Holland, Netherlands
         \and
         University of Rochester, Rochester, NY, USA
         \and
         Netherlands Organization for Applied Scientific Research (TNO), 2628 CK Delft, The Netherlands
         \and
         Max-Planck Institut f\"ur Extraterrestrische Physik, Garching bei München, Germany 
         \and
         University of Toledo, Toledo, OH, USA
         \and
         University of Massachusetts Amherst, Amherst, MA, USA
         \and
         Department of Astronomy and Astrophysics Tata Institute of Fundamental Research, Homi Bhabha Road, Colaba, Mumbai 400005, India
         \and
         Academia Sinica Institute of Astronomy \& Astrophysics, 11F of Astro-Math Bldg., No.1, Sec. 4, Roosevelt Rd., Taipei 10617, Taiwan, R.O.C.
         \and
        Department of Astronomy, The University of Texas at Austin, 2515 Speedway, Stop C1400, Austin, Texas 78712-1205, USA
         \and
         Space Telescope Science Institute, 3700 San Martin Drive, Baltimore, MD 21218, USA
         \and
       Instituto de Astrof{\'i}sica de Andaluc{\'i}a, CSIC, Glorieta de la Astronom{\'i}a s/n, E-18008 Granada, Spain
         \and
       Max Planck Institute for Astronomy, Heidelberg, Baden Wuerttemberg, Germany
        \and
        United Kingdom Astronomy Technology Centre, Edinburgh, UK
        \and
        Department of Astronomy, University of Illinois, 1002 West Green St, Urbana, IL 61801, USA
        \and
      RIKEN Cluster for Pioneering Research, Wako-shi, Saitama, 351-0198, Japan
      %   \and
      % Ritter Astrophysical Research Center, Dept. of Physics and Astronomy, University of Toledo, Toledo, OH, USA
      %    \and
      %  SKA Observatory, Jodrell Bank, Lower Withington, Macclesfield SK11 9FT, UK
      %    \and
      %  INAF-Osservatorio Astronomico di Capodimonte, Italy
      %    \and
      %  Space Telescope Science Institute, 3700 San Martin Drive, Baltimore, MD 21218, USA
      %    \and
      %  Caltech/IPAC, Pasadena, CA, USA
      %    \and
      %  Jet Propulsion Laboratory, Pasadena, CA, USA
      %   \and
      % University of Michigan, Ann Arbor, MI, USA
      %   \and
      % Space Science Institute, Boulder, CO, USA
      %   \and
      %   Center for Astrophysics Harvard \& Smithsonian, Cambridge, MA, USA
      %   \and
      % Friedrich-Schiller-Universit\"at, Jena, Th\"uringen, Germany
      %   \and
      % Department of Astronomy, University of Illinois, 1002 West Green St, Urbana, IL 61801, USA
      %   \and
      % National Radio Astronomy Observatory, 520 Edgemont Rd., Charlottesville, VA 22903 USA
      %   \and
      % Gemini South Observatory, La Serena, Chile
      %   \and
      % Departamento de Astronom\'{i}a, Universidad de Concepci\'{o}n,Casilla 160-C, Concepci\'{o}n, Chile
      %   \and
      % European Southern Observatory, Garching bei M\"unchen, Germany 
             }

   \date{Received 22 November 2023 / Accepted 15 January 2024}

% \abstract{}{}{}{}{} 
% 5 {} token are mandatory
 
  \abstract{Nitrogen-bearing complex organic molecules have been commonly detected in the gas phase but not yet in interstellar ices. This has led to the long-standing question of whether these molecules form in the gas phase or in ices. \textit{James Webb} Space Telescope (\textit{JWST}) offers the sensitivity, spectral resolution, and wavelength coverage needed to detect them in ices and investigate whether their abundance ratios are similar in gas and ice. We report the first tentative detection of CH$_3$CN, C$_2$H$_5$CN, and the simple molecule, N$_2$O, based on the CN-stretch band in interstellar ices toward three (HOPS 153, HOPS 370, and IRAS 20126+4104) out of the five protostellar systems observed as part of the Investigating Protostellar Accretion (IPA) GO program with \textit{JWST}-NIRSpec. We also provide upper limits for the two other sources with smaller luminosities in the sample. We detect OCN$^-$ in the ices of all sources with typical CH$_3$CN/OCN$^-$ ratios of around 1. Ice and gas column density ratios of the nitrogen-bearing species with respect to each other are better matched than those with respect to methanol, which are a factor of ${\sim}5$ larger in the ices than the gas. We attribute the elevated ice column densities with respect to methanol to the difference in snowline locations of nitrogen-bearing molecules and of methanol, biasing the gas-phase observations toward fewer nitrogen-bearing molecules. Moreover, we find tentative evidence for enhancement of OCN$^-$, CH$_3$CN, and C$_2$H$_5$CN in warmer ices, although formation of these molecules likely starts along with methanol in the cold prestellar phase. Future surveys combining NIRSpec and MIRI, and additional laboratory spectroscopic measurements of C$_2$H$_5$CN ice, are necessary for robust detection and conclusions on the formation history of complex cyanides.}
   %{context}
  % {aims}
  % {methods}
  % {results}
  % {conclusions}

%We also report detection of icy OCN$^{-}$ in the five protostellar envelopes. Ice column density ratios of OCN$^{-}$ to CH$_3$OH (a molecule that is accepted to be formed in ices) are higher than the gas-phase column density ratios of HNCO/CH$_3$OH by a factor of ${\sim}5$. This suggests that  

   \keywords{Astrochemistry --
                Stars: Low-mass --
                Stars: protostars --
                ISM: abundances --
                ISM: molecules --
                Techniques: spectroscopic
               }

   \maketitle
%
%-------------------------------------------------------------------

\section{Introduction}
\label{sec:intro}

Complex organic molecules (COM; species with $\geq6$ atoms containing carbon) are thought to be the precursors of larger species that are necessary for habitable worlds (\citealt{Herbst2009}). COMs have been observed in the gas phase in various stages of star formation such as the prestellar phase (e.g., \citealt{Bacmann2012}; \citealt{JimenezSerra2016}), protostellar phase (e.g., \citealt{vanDishoeck1995}; \citealt{Cazaux2003}; \citealt{Jorgensen2016}; \citealt{Lee2017}; \citealt{Belloche2020}; \citealt{vanGelder2020}; \citealt{MartinDomenech2021}; \citealt{Nazari2021}; \citealt{Diaz2022}) and even protoplanetary disks (e.g., \citealt{Oberg2015}; \citealt{Brunken2022}). These observations have motivated many laboratory and modeling works to understand how and when COMs form (see reviews by \citealt{Jorgensen2020}; \citealt{Ceccarelli2023}). 

Multiple laboratory studies argue for the formation of these species in icy dust grain mantles in either the translucent phase, cold prestellar phase or during the later warm-up phase around the forming protostar (\citealt{Oberg2009}; \citealt{Meinert2016}; \citealt{Fedoseev2017}; \citealt{Munoz2019}; \citealt{Qasim2019}; \citealt{Chuang2021}; \citealt{Ioppolo2021}; \citealt{Santos2022}). If formed in ices, COMs may now be directly detected in protostellar ices with the \textit{James Webb} Space Telescope (\textit{JWST}), while the telescopes preceding \textit{JWST}, such as the \textit{Infrared Space Observatory} (\textit{ISO}) and the \textit{Spitzer Space Telescope}, did not have the sensitivity, spectral resolution and wavelength coverage (particularly ${\sim}4.4$\,$\mu$m) required to securely detect COMs. Although those telescopes provided valuable information on simple species such as H$_2$O, CO$_2$, CH$_4$ and NH$_3$, they could only provide upper limits on most COMs, except for methanol (CH$_3$OH) which was detected (\citealt{Grim1991}; \citealt{Gerakines1999}; \citealt{Boogert2000}; \citealt{Keane2001}; \citealt{Gibb2004}; \citealt{Oberg2008}; \citealt{Pontoppidan2008}; \citealt{Bottinelli2010}; \citealt{Oberg2011}; \citealt{Boogert2015}). 

If the icy grains are heated, COMs can sublimate into the gas close to the protostar and become observable to telescopes such as the Atacama Large Millimeter/submillimeter Array (ALMA) and the Northern Extended Millimeter Array (NOEMA). Once ices sublimate, gas-phase chemistry may occur to either form (larger) COMs from simple species or to alter the abundances of the sublimated COMs (\citealt{Charnley1992}; \citealt{Herbst2009}; \citealt{Barone2015}; \citealt{Balucani2015}; \citealt{Vazart2020}). Many chemical models have been developed to explain formation of these species on the dust grains or in the gas phase (\citealt{Garrod2008}; \citealt{Aikawa2008}; \citealt{Walsh2014}; \citealt{Garrod2022}).

Recent ALMA and NOEMA surveys have provided valuable information on the gas-phase column density ratios of COMs in a large number of protostellar systems (e.g., \citealt{Belloche2020}; \citealt{Yang2021}; \citealt{Nazari2022ALMAGAL}; \citealt{Chen2023Cocoa}; \citealt{Taniguchi2023}). These works frequently show that gas-phase nitrogen- and oxygen-bearing COMs have remarkably constant column density ratios across protostellar systems with a large range of mass and luminosity (also see \citealt{Coletta2020}). These constant ratios are often interpreted as COM formation in similar environments, likely in the prestellar ices. However, the cold, hard evidence is to directly detect COMs in ices and compare the ice and gas abundances to search for any relation between the two. 

%A similar (within factor of ${\sim}2-3$) ice and gas abundance ratio would then point to the bulk formation of COMs in ices and their intact sublimation from those ices into the gas, while a large difference between the gas and ice abundance ratios could suggest additional gas-phase chemistry, vastly different initial conditions or simply physical effects such as differences in sublimation temperatures   

With the launch of \textit{JWST}, COMs in ices can be studied with the spectroscopic instruments; Mid-InfraRed Instrument (MIRI) and Near InfraRed Spectrograph (NIRSpec). Already, hints for oxygen-bearing COMs are observed in MIRI studies (\citealt{Yang2022}; \citealt{Rocha2023}) building on the \textit{ISO} and \textit{Spitzer} results (\citealt{Schutte1999}; \citealt{Oberg2011}). Nitrogen-bearing COMs are as important, in particular, because nitrogen contributes to forming amino acids and nucleobases required to develop habitable worlds. Yet, elemental nitrogen is a factor of ${\sim}5$ less abundant than oxygen, making it more difficult to observe N-bearing molecules. Therefore, much less is known about the formation and evolution of N-bearing COMs in ices or in the gas. In fact, the only N-bearing COM considered in interstellar ices with \textit{JWST} is formamide (NH$_2$CHO) where \cite{Slavicinska2023NH2CHO} used the Early Release Science program, Ice Age (\citealt{McClure2023}) and the guaranteed time observation program, JOYS (\citealt{Ewine2023}) to hunt for formamide at 7.24\,$\mu$m, which was suggested previously to be in protostellar ices by \cite{Schutte1999}. Even then, \cite{Slavicinska2023NH2CHO} could find only upper limits on formamide column density. In this paper we use \textit{JWST}-NIRSpec to search for nitrogen-bearing COMs through their strong CN stretching band and provide the first tentative detection of N-bearing COMs in ices. 

We focus on methyl cyanide (CH$_3$CN) and ethyl cyanide (C$_2$H$_5$CN). They have a strong vibrational transition in the ${\sim}4.4-4.7$\,$\mu$m range. Although methyl cyanide is one of the most used tracers of the hot core phase, its formation mechanism has been debated for decades (\citealt{Huntress1979}; \citealt{Wilner1994}; \citealt{Walsh2014}; \citealt{LeGal2019}; \citealt{Nazari2023CGD}). Moreover, it is the most abundant nitrogen-bearing COM in the gas-phase (e.g., \citealt{Calcutt2018}; \citealt{Yang2021}). Therefore, if the bulk of CH$_3$CN were to form in ices, it is one of the few nitrogen-bearing species that could be observed by \textit{JWST} providing important clues to its formation mechanism. Previously, \cite{Rachid2022} investigated the presence of CH$_3$CN in \textit{ISO} (also see \citealt{Gibb2004}) and \textit{Spitzer} spectra and found upper limits on its ice column densities. We also include C$_2$H$_5$CN in our work given that it has a transition right next to CH$_3$CN, so it is impossible to quantify one without the other. Furthermore, the chemistry of C$_2$H$_5$CN is thought to be related to CH$_3$CN (\citealt{Bulak2021}), and it has high gas-phase abundances with C$_2$H$_5$CN/CH$_3$CN${\sim} 0.1$ (\citealt{Nazari2022ALMAGAL}). The hunt for CH$_3$CN and C$_2$H$_5$CN is now facilitated by the recent laboratory studies on their spectroscopy (\citealt{Moore2010}; \citealt{Rachid2022}). 

In addition, we use the column density of OCN$^-$ (${\sim}4.6$\,$\mu$m; \citealt{Broekhuizen2005}) in ices as a reference species. Given that OCN$^-$ is thought to form from HNCO (a molecule observed abundantly in the gas phase) and NH$_3$ in ices (\citealt{Schutte1999}; \citealt{Novozamsky2001}; \citealt{Schutte2003}; \citealt{Ruaud2016}), OCN$^-$ is an ideal reference for nitrogen-bearing molecules. When ices sublimate close to the protostar, OCN$^-$ is thought to turn back into HNCO as it sublimates into the gas (\citealt{Oberg2009HNCO}). Therefore, we compare the ice column density ratios with respect to OCN$^-$ with the gas column density ratios with respect to HNCO.   

In this work we search for CH$_3$CN and C$_2$H$_5$CN around the protostars observed by \textit{JWST} as part of the Investigating Protostellar Accretion (IPA; \citealt{Federman2023}; \citealt{Narang2023}) program and tentatively detect them around three protostars for the first time. We first fit a local continuum to these data, fit laboratory data to the spectra and measure the CH$_3$CN, C$_2$H$_5$CN and OCN$^-$ ice column densities (Sect. \ref{sec:columns}). We then compare our column density ratios with the gas-phase ratios in Sect. \ref{sec:comp_gas}, discuss the chemical interpretations of our findings in Sect. \ref{sec:chem} and conclude in Sect. \ref{sec:conclusion}.

\begin{figure*}
    \centering
    \includegraphics[width=0.9\textwidth]{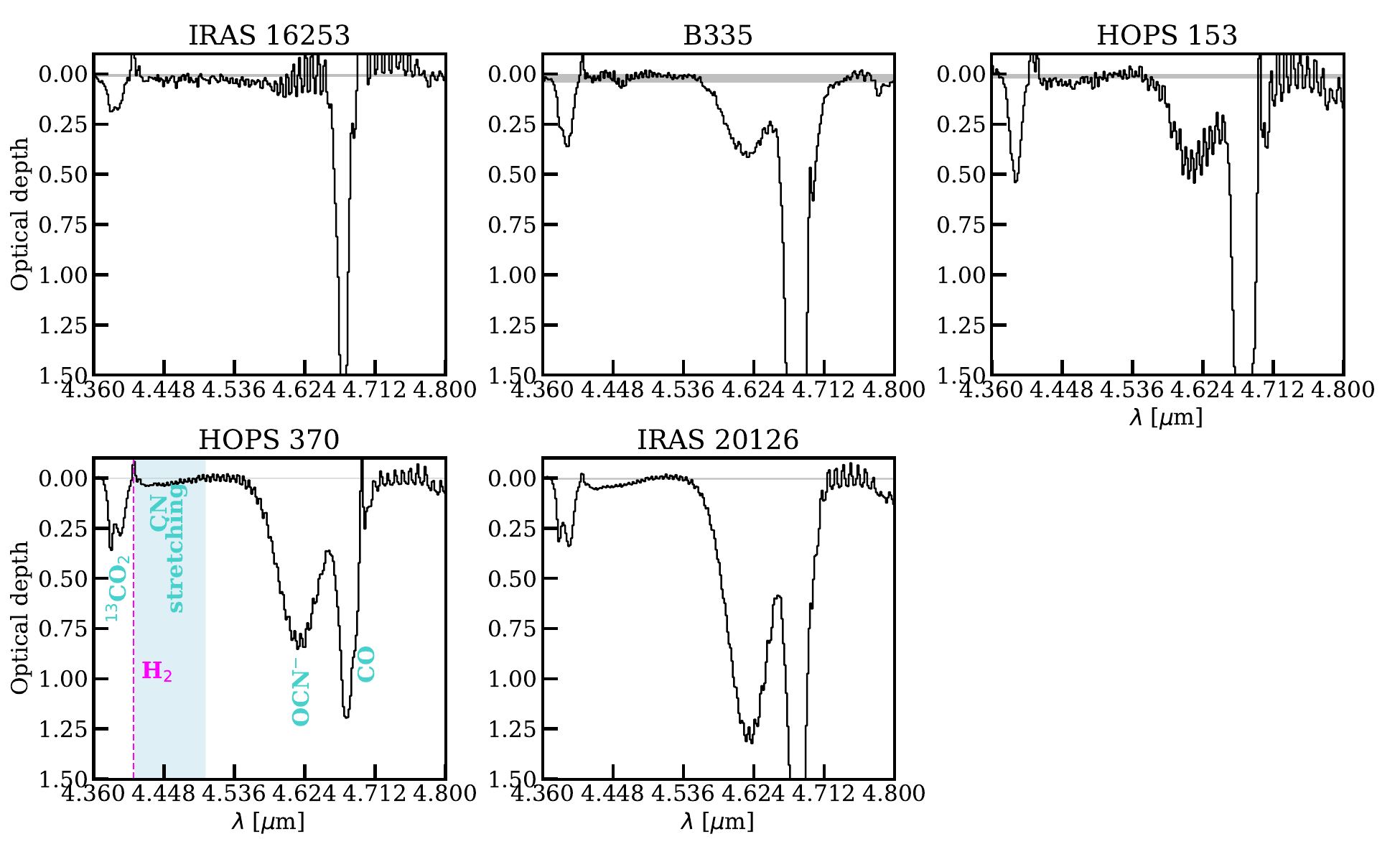}
    \caption{Calculated optical depth for all 5 sources in the wavelength range of interest. Gray shaded area shows the optical depth $3\sigma$ uncertainty level. Although the spectra are extracted off source (see table \ref{tab:obs}), some CO emission lines are still seen in the spectra. The CN stretching mode vibration band is shaded with light blue. The emission line between the $^{13}$CO$_2$ ice feature and the CN stretching band is the H$_2$ $0-0$ S(10) line (marked with the dashed magenta line).} 
    \label{fig:optical_depth}
\end{figure*}

\section{Observations and methods}
\subsection{Data}
\label{sec:data}

Here we study the five protostars observed by \textit{JWST} NIRSpec (\citealt{Bagnasco2007}; \citealt{Jakobsen2022}; \citealt{Boker2023}) in the Integral Field Spectroscopy mode (IFU; \citealt{Boker2022}) as part of the IPA program (project ID: 1802, PI: T. Megeath). The protostellar systems observed by this program have a range of bolometric luminosities (${\sim}0.2-10^4$\,L$_{\odot}$) and protostellar masses (${\sim}0.12-12$\,M$_{\odot}$). They are, in order of increasing luminosity, IRAS 16253-2429 (hereafter IRAS 16253), B335-IRS (hereafter B335), HOPS 153, HOPS 370, and IRAS 20126+4104 (hereafter IRAS 20126). Table \ref{tab:obs} presents the bolometric luminosities and distances of the objects. Data reduction is explained in detail in \cite{Federman2023} and \cite{Narang2023}; thus we give only a brief overview of the data. The objects were observed in the G395M setting (spectral resolving power $R$ of ${\sim}1000$) with a 4-point dither pattern, a spatial coverage of ${\sim} 6\arcsec \times 6\arcsec$, and a spatial resolution of ${\sim}0.2\arcsec$. The observations cover the wavelength range between ${\sim}$2.87\,$\mu$m and ${\sim}$5.27\,$\mu$m. This paper focuses on the wavelength range of ${\sim}4.4-4.7$\,$\mu$m, as complex and simple species with a CN bond such as CH$_3$CN, C$_2$H$_5$CN, HCN, HC$_3$N, and OCN$^-$ are all known to have one of their strongest transitions (the CN stretch) around this region (\citealt{Soifer1979}; \citealt{Russo1996}; \citealt{Gibb2000}; \citealt{Moore2010}; \citealt{Rachid2022}).

This work focuses on the ice features that are absorbed in the continuum, but the wavelength range of interest is also where CO rotational-vibrational gas-phase emission lines are known to lie which make the analysis of the ices more difficult (\citealt{Federman2023}; \citealt{Rubinstein2023}). Therefore, we avoided these emission lines in two ways. First, we searched the data cubes for the best spatial regions at which these gas-phase lines are less prominent. For each source we extracted the spectrum from such a region where CO emission lines are either not observed or are relatively weak. However, due to the potential remaining uncertainty by the residual gas-phase CO lines, we also analyzed the spectra extracted from the same positions using the emission-line subtracted cubes produced by \cite{Rubinstein2023}. This method provided two sets of spectra with either no gas-phase emission lines or relatively weak ones which facilitated the identification of the CN stretching band. 

In all cases the spectra were extracted off-source with their centers given in Table \ref{tab:obs} (also see Fig. \ref{fig:extract}). The radius of the aperture ($0.6\arcsec$) is taken as ${\sim}3$ times the full width at half maximum (FWHM) of the point spread function of NIRSpec-IFU at 4.9\,$\mu$m. The rms on the continuum optical depth was calculated using the error plane of each data cube by selecting two spectral regions with least (or no) spectral features. The $\sigma$ on optical depth ranges from ${\sim}0.001$ to ${\sim}0.01$ (see Table \ref{tab:obs}). 

\begin{figure*}
    \centering
    \includegraphics[width=17cm]{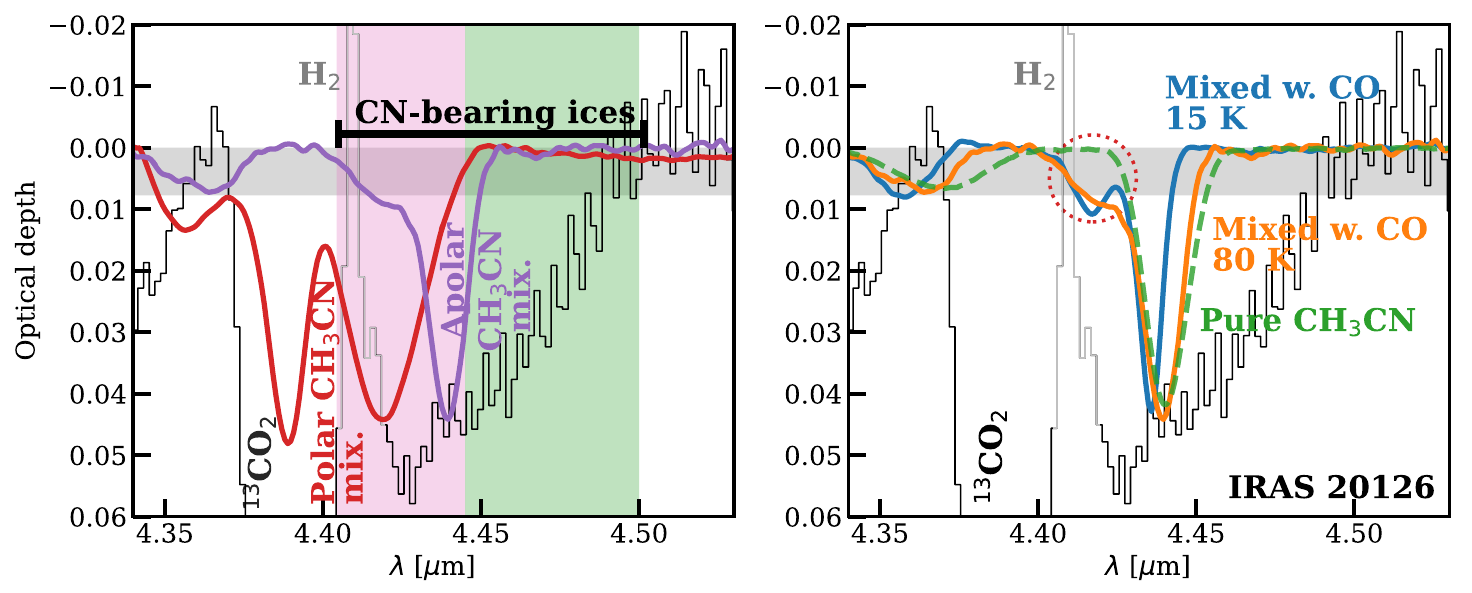}
    \caption{Continuum subtracted spectrum of IRAS 20126 in the $4.35-4.5$\,$\mu$m range. The shaded gray area shows the $3\sigma$ uncertainty on optical depth. The H$_2$ emission line at ${\sim}4.41$\,$\mu$m is shown in gray. \textit{Left}: Two of the most extreme CH$_3$CN ice mixtures in peak position with large FWHM from \cite{Rachid2022} are shown in red and purple to demonstrate that the shaded green region cannot be explained by CH$_3$CN mixtures but the shaded pink region can. Red is the CH$_3$CN:H$_2$O:CO$_2$ (1:5:2, 50\,K) mixture and purple is the CH$_3$CN:CO (1:10, 80\,K). \textit{Right}: mixures of CH$_3$CN:CO (1:10) at 15\,K (blue) and 80\,K (orange) to demonstrate that the one at 80\,K has a larger FWHM and peak position. In addition, the red circle shows the shoulder that does not exist in the pure CH$_3$CN (dashed green) laboratory spectra of \cite{Rachid2022} and is produced in the CH$_3$CN ice feature when mixed with CO.}
    \label{fig:demo}
\end{figure*}

%to include all the relevant flux

\subsection{Continuum determination and subtraction}

The continuum was fitted locally and subsequently subtracted using the polynomial function method included in the ENIIGMA fitting tool (\citealt{Rocha2021}) to isolate the bands of interest from the broad water combination mode at ${\sim}4.45$\,$\mu$m. This continuum identification method has been commonly used in the literature for ice analysis (e.g., \citealt{Oberg2011}; \citealt{Boogert2022}; \citealt{Slavicinska2023NH2CHO}). Figure \ref{fig:cont} presents the continuum fitted to the data. We note that grain shape effects are not expected to dominate and change the ice features in our region of interest (e.g., \citealt{Palumbo1995}; \citealt{Dartois2022}). The continuum subtracted fluxes ($F_{\lambda}$) were then converted to optical depth through $-\ln{(F^{\rm data}_{\lambda}/F^{\rm cont}_{\lambda})}$. Figure \ref{fig:optical_depth} presents the continuum-subtracted spectra converted into the optical depth scale for the five sources. The $3\sigma$ on optical depth is highlighted on Fig. \ref{fig:optical_depth}; it is larger for the lower luminosity sources. 

The three most luminous sources (HOPS 153, HOPS 370, and IRAS 20126) show a significant ($S/N{\gtrsim}3$) absorption feature at $4.44$\,$\mu$m. Continuum identification for IRAS 16253 is uncertain due to the remaining CO emission lines, hence the significance of the absorption feature at $4.44$\,$\mu$m is unclear. For B335 the absorption feature is at around the $3\sigma$ level but because of the residual CO emission lines, it is difficult to conclude. Using the emission subtracted spectra the absorption feature of B335 is less than the $3\sigma$ level and thus is not significant. Hence, we focus our analysis only on the three most luminous sources.

%Moreover, for the locations at which the spectra are extracted, there are some variations in the strength of the OCN$^-$ band at $4.62$\,$\mu$m across the different protostellar systems. 

%For example, it is prominently detected in HOPS 370 and IRAS 20126 while at best an upper limit can be found on its column density for IRAS 16253. 

%This figure also indicates the regions used in the spectrum to determine the local continuum with red circles.  

%This is to minimize the effect of CO emission lines on the absorption feature in the ${\sim}4.4-4.52$\,$\mu$m range. 

\section{Band identification and fitting}
\label{sec:ident}
\subsection{CH$_3$CN}

Figure \ref{fig:demo} presents a blow-up of the 4.35-4.5\,$\mu$m range for IRAS 20126, which includes the the $^{13}$CO$_2$ ice feature at $\leq 4.40$\,$\mu$m and the CN-bearing species absorption feature at the longer wavelengths ($4.4-4.5$\,$\mu$m). \cite{Rachid2022} considered CH$_3$CN infrared spectra in seven different ice mixtures and various temperatures (15-140\,K). The line profiles of these various mixtures peak between 4.410\,$\mu$m up to 4.442\,$\mu$m. The left panel of Fig. \ref{fig:demo} presents two of the CH$_3$CN ice mixture spectra from \cite{Rachid2022} that have the shortest and longest peak wavelengths as well as a large FWHM. Figure \ref{fig:demo} shows that although CH$_3$CN could be responsible for the absorption feature at $4.4-4.45$\,$\mu$m, the entire feature up to 4.5\,$\mu$m is too wide to be explained by CH$_3$CN mixtures alone, thus Sect. \ref{sec:ident_C2H5CN} explores other species that could be responsible for the rest of the absorption feature.

%After this initial exploration we opted for fitting the ${\sim}4.4-4.52$\,$\mu$m band with a minimum number of combination of laboratory measurements of CH$_3$CN, C$_2$H$_5$CN, and N$_2$O. 

We opted for fitting the ${\sim}4.40-4.45$\,$\mu$m region with a minimum number of CH$_3$CN ice mixtures. We used Fig. 3 (FWHM versus peak position) and Table B.2 of \cite{Rachid2022} to carefully determine the ice mixtures that best fit the data. The minimum number of components requires features with large FWHM. The mixtures from \cite{Rachid2022} that match the absorption band at $\lambda>4.4$\,$\mu$m in peak position and have relatively large FWHM (${\sim}8-15$\,cm$^{-1}$) are CH$_3$CN:CO$_2$, CH$_3$CN:NH$_3$, CH$_3$CN:CO (at higher temperatures) and mixtures with water. From these mixtures, those with water (polar ices) have the largest FWHM, and from those, the mixture CH$_3$CN:H$_2$O:CO$_2$ is favored to CH$_3$CN:H$_2$O and CH$_3$CN:H$_2$O:CH$_4$:NH$_3$ because it peaks at longer wavelengths. Although favored, the mixture of CH$_3$CN:H$_2$O:CO$_2$ cannot fit the entire absorption feature thus, at least one additional component is needed to fit the ${\sim}4.40-4.45$\,$\mu$m feature. 

%because it peaks at around where the H$_2$ S(10) emission line falls, and absorption would remain unfitted between the peak wavelength (4.419\,$\mu$m) and 4.44\,$\mu$m (see left panel of Fig. \ref{fig:demo}). Therefore,

\begin{figure*}
    \centering
    \includegraphics[width=\textwidth]{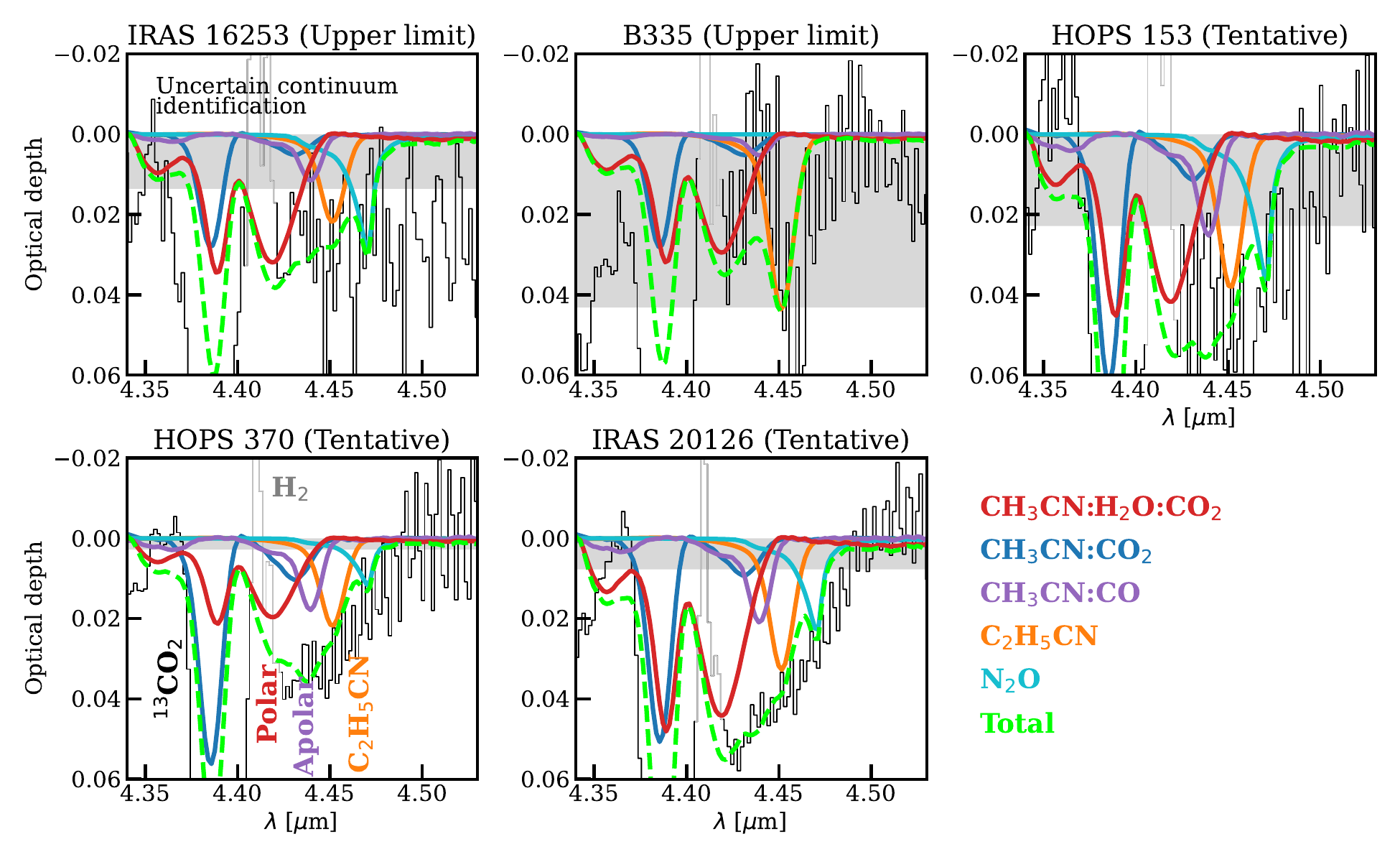}
    \caption{Fitted CH$_3$CN, C$_2$H$_5$CN and N$_2$O models on top of the continuum subtracted data (black). The models of the various mixtures of the ice are color coded as follows, CH$_3$CN:H$_2$O:CO$_2$ (1:5:2, 50\,K, red), CH$_3$CN:CO$_2$ (1:10, 15\,K, blue), CH$_3$CN:CO (1:10, 80\,K, purple), C$_2$H$_5$CN (orange, 50\,K), and N$_2$O (70\,K, cyan). The total fit is shown with the dashed green line. The $3\sigma$ uncertainty on optical depth is shown in shaded gray. Continuum identification of IRAS 16253 is uncertain and thus the significance of the observed feature is unclear. The H$_2$ emission line is shown in gray and ignored when fitting (see also Fig. \ref{fig:fit_baseline}).} 
    \label{fig:fits}
\end{figure*}

We chose the apolar mixture CH$_3$CN:CO as this additional component. Compared to CH$_3$CN:CO$_2$, CH$_3$CN:CO at high temperatures peaks at longer wavelengths, so it better matches the data. Between the mixtures with CO or NH$_3$ we opted for CH$_3$CN:CO for two reasons: (i) the mixture with CO has a shoulder at shorter wavelengths (${\sim}4.41$\,$\mu$m, see right panel of Fig. \ref{fig:demo}) that avoids adding unnecessary additional components, and (ii) the CO mixture is likely more relevant for interstellar ices (\citealt{Boogert2015}). With CH$_3$CN:H$_2$O:CO$_2$ and CH$_3$CN:CO the main absorption feature can be reasonably fitted, but we included the CH$_3$CN:CO$_2$ mixture to avoid having a dip in between the peak of CH$_3$CN:H$_2$O:CO$_2$ and CH$_3$CN:CO. Given the present uncertainties it is not possible to argue whether CH$_3$CN:CO$_2$ is strictly necessary to fit the absorption feature. We emphasize that our final fits (Fig. \ref{fig:fits}) are not unique and a fit can also be obtained by for example increasing the CH$_3$CN:CO$_2$ component and decreasing the CH$_3$CN:H$_2$O:CO$_2$ component. However the final column density of CH$_3$CN will only be affected by less than a factor of ${\sim}2$.

It is not possible to constrain the temperature of the ice for CH$_3$CN:CO$_2$ and CH$_3$CN:H$_2$O:CO$_2$ mixtures apart from arguing that they should be below ${\sim}50$\,K and ${\sim}120$\,K under laboratory conditions, respectively, to retain the large FWHM. However, for the CH$_3$CN:CO mixture, high temperatures (${\sim}80$\,K as opposed to ${\sim}15$\,K) better match the data by producing a larger FWHM and being centered at longer wavelengths (see right panel of Fig. \ref{fig:demo}). We note that ${\sim}80$\,K in the laboratory conditions is ${\sim}30$\,K in interstellar conditions (e.g., \citealt{Boogert2015}). This implies that the temperatures of ice mixtures under laboratory conditions are not equivalent to the true ice temperature in the protostellar environments because of the different time scales in those two conditions. In terms of ice morphology, this means that phase transitions and structural changes (e.g., amorphous to crystalline), would happen at higher temperatures in the laboratory. Thus, the 80\,K ice mixture of CH$_3$CN and CO in the laboratory which still includes CO trapped in the ice is similar to a 30\,K ice mixture of CH$_3$CN and CO in protostellar environments with most CO likely sublimated, but some trapped in the ice. This trapping can be seen from the effect of CO on the CH$_3$CN absorption feature at ${\sim}4.41$\,$\mu$m (see the difference between CH$_3$CN:CO mixture and pure CH$_3$CN in right panel of Fig. \ref{fig:demo}).   

%We use Gaussians with the parameters given in \cite{Rachid2022} for the final chosen mixtures to fit the data.

\subsection{C$_2$H$_5$CN and N$_2$O}
\label{sec:ident_C2H5CN}

There is considerable absorption (at ${\gtrsim}3\sigma$) in the $4.442 < \lambda < 4.5$\,$\mu$m region of IRAS 20126 (see left panel of Fig. \ref{fig:demo}). Therefore, we searched the literature for other simple and complex molecules with absorption bands in this region. In total we searched for ${\sim}30$ pure ices or ice mixtures at various temperatures (${\sim}10-150$\,K) to examine potential contribution from those in the ${\sim}4.4-4.52$\,$\mu$m region. We particularly used NASA's Optical Constants database\footnote{https://ocdb.smce.nasa.gov/search/ice?t=gw9anjm2av}, the Cosmic Ice Laboratory database\footnote{https://science.gsfc.nasa.gov/691/cosmicice/constants.html} and the Leiden Ice Database for Astrochemistry (\citealt{Rocha2022}). This search included Oxygen-bearing species such as CH$_3$OCH$_3$, CH$_3$CHO and CO in addition to many nitrogen-bearing species such as HNCO and N$_2$O, and in particular those with a CN bond, C$_2$H$_5$CN, C$_2$H$_3$CN, C$_2$N$_2$, and HC$_{n}$N. Among the considered species, C$_2$H$_5$CN and N$_2$O in addition to CH$_3$CN were found to best match the absorption feature observed at ${\sim}4.4-4.52$\,$\mu$m. The other species did not match the data because either their peak did not match or if they had an absorption feature at the right wavelength, they showed a stronger one at a wavelength with no absorption feature in the spectrum (e.g., HCN).

\begin{figure*}
    \centering
    \includegraphics[width=0.9\textwidth]{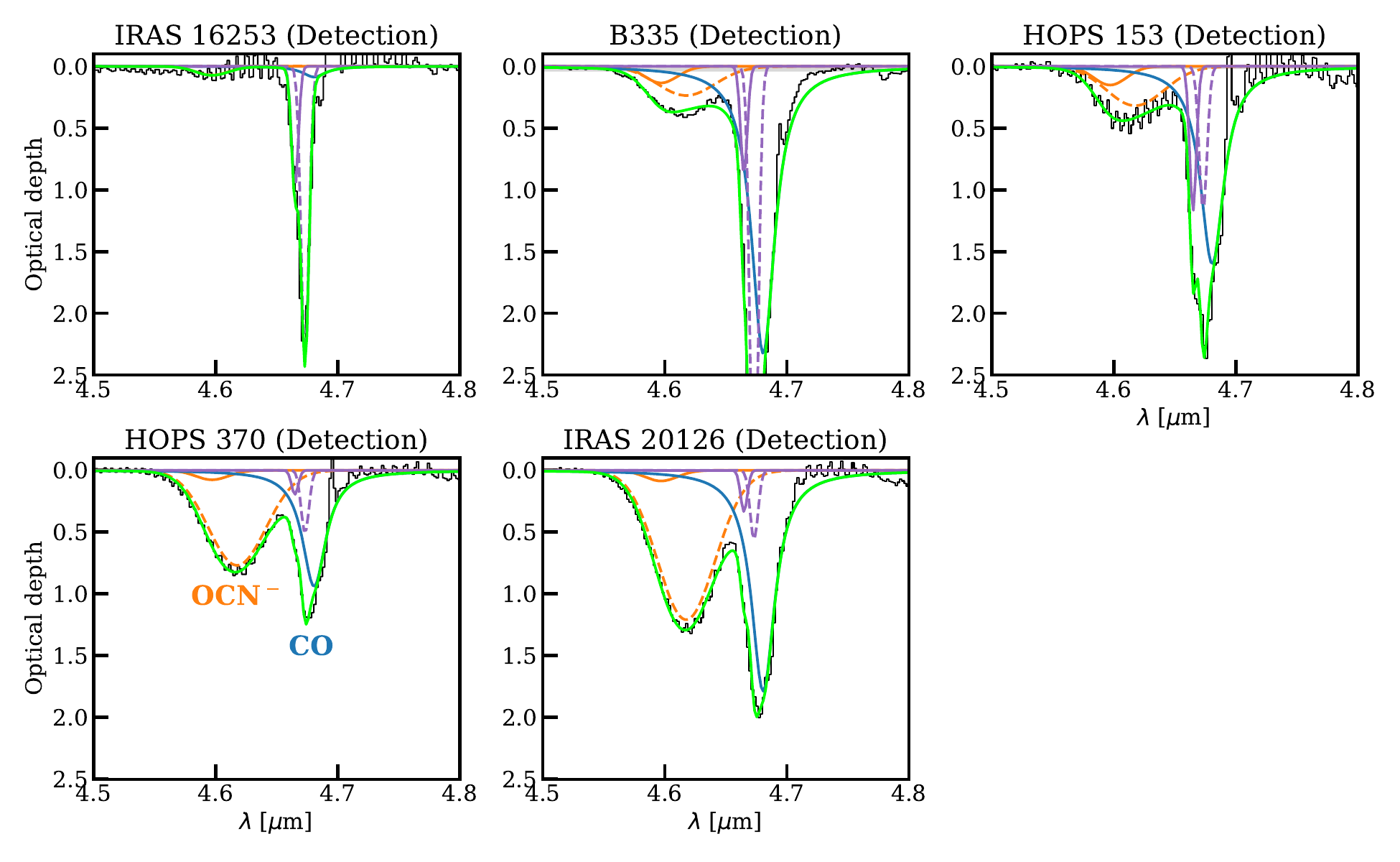}
    \caption{Fitted model to OCN$^-$ absorption feature. Orange dashed and solid lines show the red and blue OCN$^{-}$ (sometimes attributed to CO absorption instead), showing that the bulk of the XCN feature is dominated by OCN$^-$. The blue and purple lines show the Lorentzian and Gaussian contributions to the CO band, respectively.The XCN and CO bands are fitted simultaneously. The total fit is shown in green. The $3\sigma$ uncertainty on optical depth is shown in shaded gray. The y-axis limit is set for better visibility of the ice features in all sources which cuts off the fit to the CO band in B335. The fit to the CO ice band in B335 is similar to the other sources and does not over predict the data.} 
    \label{fig:OCN_fit}
\end{figure*}

To the best of our knowledge, there is no laboratory study of ethyl cyanide or N$_2$O ice mixtures at infrared wavelengths. Therefore, we took the pure amorphous and crystalline absorbance spectrum of N$_2$O at 10\,K and 70\,K from \cite{Gerakines2020}. The 70\,K spectrum better fit the data because its peak is at shorter wavelengths. Moreover, we took the imaginary refractive indices ($k_{\nu}$) of pure laboratory measurements of amorphous C$_2$H$_5$CN at various temperatures (50-110\,K) from \cite{Moore2010}. Before fitting the observed spectra we converted $k_{\nu}$ from \cite{Moore2010} to optical depth using

\begin{equation}
    \tau = 4 \pi k_{\nu} \nu d,
\end{equation}

\noindent where $\nu$ is the wavenumber and $d$ is the ice thickness of 2.29\,$\mu$m from the experiments of \cite{Moore2010}. This is a first order approximation because we do not have the information on the Fresnel coefficients for these measurements.

Figure \ref{fig:fits} presents our final model fits to the $4.4-4.52$\,$\mu$m band and the decomposition of that model into components for the five objects. We also fitted the spectra with gas-phase emission lines subtracted and the final models for those are shown in Fig. \ref{fig:fit_baseline}. Using either the spectra before or after subtraction of emission lines we obtain only upper limits on CH$_3$CN and C$_2$H$_5$CN toward IRAS 16253 (due to the uncertain continuum identification) and B335. However, using both versions of the spectra we tentatively detect CH$_3$CN, C$_2$H$_5$CN, and N$_2$O toward the other three sources, all for the first time in interstellar ices. 

\subsection{OCN$^-$}

The absorption feature at 4.61\,$\mu$m is often referred to as the `XCN' band and is extensively studied (\citealt{Gibb2004}; \citealt{Broekhuizen2005}; \citealt{Fraser2005}; \citealt{Oberg2011}; \citealt{Mate2012}; \citealt{Boogert2022}). This band was suggested by \cite{Broekhuizen2005} to be composed of two Gaussian profiles (at 2165.7\,cm$^{-1}$ with FWHM of 26\,cm$^{-1}$ and at 2175.4\,cm$^{-1}$ with FWHM of 15\,cm$^{-1}$). One (2165.7\,cm$^{-1}$) can be attributed to OCN$^-$ and the other one has been attributed to either CO bonding to grain surfaces or OCN$^-$ in apolar ice environments (\citealt{Fraser2005}; \citealt{Oberg2011}). Moreover, the XCN band is blended with the stretching mode of CO which will affect the fitting of this region. The CO band is proposed to be composed of a Lorentzian profile (peak at 2136.5\,cm$^{-1}$ with FWHM 10.6\,cm$^{-1}$) and two Gaussian profiles (peaks at 2139.9\,cm$^{-1}$ and 2143.7\,cm$^{-1}$ with FWHM of 3.5\,cm$^{-1}$ and 3.0\,cm$^{-1}$) by \cite{Pontoppidan2003}. Therefore, we use these line profiles in addition to the two Gaussian profiles suggested for the XCN band by \cite{Broekhuizen2005} to fit the feature between 4.56\,$\mu$m and 4.7\,$\mu$m simultaneously. This method has also been used in \cite{Boogert2022} for several high-mass protostars. The fits, which match the data very well, are shown in Fig. \ref{fig:OCN_fit}.

\section{Column densities}
\label{sec:columns}

\subsection{CH$_3$CN and C$_2$H$_5$CN}

Column densities for CH$_3$CN and C$_2$H$_5$CN are calculated using, 

\begin{equation}
    N = \int{\tau d\nu}/A,
    \label{eq:column}
\end{equation}

\noindent where $A$ is the band strength, $\tau$ is the optical depth and the integral is performed in the wave number space. The measured ice column densities or upper limits are given in Table \ref{tab:results}. The band strengths ($A$) used for determination of ice column densities are given in Table \ref{tab:band_strengths}. The band strengths for the CH$_3$CN ice mixtures are taken from \cite{Rachid2022} as 0.8, 0.8, and 1.75 times the band strength of pure CH$_3$CN (1.9$\times 10^{-18}$\,cm\,molecule$^{-1}$) for CH$_3$CN:CO$_2$, CH$_3$CN:CO and CH$_3$CN:H$_2$O:CO$_2$, respectively. The band strength for C$_2$H$_5$CN (50\,K) is calculated as $2.86 \times 10^{-18}$\,cm\,molecule$^{-1}$ using the following equation (e.g., \citealt{Roser2021})

\begin{equation}
    A = \frac{m}{\rho N_{\rm A}} \int 4 \pi \nu k_{\nu} d\nu,
    \label{eq:A}
\end{equation}

\noindent where $m$ is the molar mass of ethyl cyanide (55.08\,g\,mol$^{-1}$), $\rho$ is its ice density (0.703\,g\,cm$^{-3}$; \citealt{Gerakines2022}), and $N_{\rm A}$ is the Avogadro’s number. The band strength of C$_2$H$_5$CN is also determined in \cite{Gerakines2022} based on the results of \cite{Moore2010} as $2.8 \times 10^{-18}$\,cm\,molecule$^{-1}$ which agrees well with what is found here. For the upper limits we take the maximum of the measured upper limit from the spectra before (Fig. \ref{fig:fits}) and after the emission line subtraction (Fig. \ref{fig:fit_baseline}). The uncertainties on column densities for the tentative detections are dominated by the error in continuum fitting and the error on band strengths. The uncertainties on the band strengths are normally on the order of ${\sim}30\%$ and those based on the choice of the continuum and the line blending are estimated as ${\sim}40\%$, resulting in a total uncertainty of around $\sqrt{0.3^2 + 0.4^2} = 50\%$ on the column densities. We note that the column densities measured from the spectra before (Table \ref{tab:results}) and after subtraction of the emission lines agree within the uncertainties.  

\begin{table}
%\Huge
\renewcommand{\arraystretch}{1.3}
    \caption{Ice column densities}
    \label{tab:results}
    \resizebox{\columnwidth}{!}{\begin{tabular}{@{\extracolsep{1mm}}*{4}{l}}
          \toprule
          \toprule      
        Sources & $N_{\rm CH_3CN}\, (\rm cm^{-2})$ &  $N_{\rm C_2H_5CN}\, (\rm cm^{-2})$ &  $N_{\rm OCN^-}\, (\rm cm^{-2})$\\
        \midrule

IRAS 16253 & $<$2.5 $\times 10^{17}$& $<$7.1 $\times 10^{16}$& 2.6 $\times 10^{16}$ \\
B335 & $<$2.0 $\times 10^{17}$& $<$1.4 $\times 10^{17}$& 5.0 $\times 10^{16}$ \\
HOPS 153 & 4.3 $\times 10^{17}$&1.2 $\times 10^{17}$& 6.8 $\times 10^{16}$ \\
HOPS 370 & 2.8 $\times 10^{17}$&7.1 $\times 10^{16}$& 1.6 $\times 10^{17}$ \\
IRAS 20126 & 4.0 $\times 10^{17}$&1.1 $\times 10^{17}$& 2.6 $\times 10^{17}$ \\

\bottomrule
        \end{tabular}}
        \tablefoot{The uncertainties are estimated at around 30\% for OCN$^-$ and at around 50$\%$ for the other two species.}
\end{table}

\subsection{OCN$^{-}$}

Here we are only interested in the OCN$^-$ column densities to be used as a reference for column density ratios because of its potential chemical relation to HNCO (\citealt{Schutte2003}; \citealt{Oberg2009HNCO}), another molecule observed abundantly in the gas-phase (e.g., \citealt{Nazari2022ALMAGAL}). Therefore, we report the column density of the red part of the XCN band (orange dashed line in Fig. \ref{fig:OCN_fit}) as the final OCN$^-$ column density. From Fig. \ref{fig:OCN_fit} it can be seen that OCN$^-$ is detected in all sources and dominates the XCN band. We emphasize that this finding is unique to the spectra extracted from the apertures off-source as described in Sect. \ref{sec:data} (also see Fig. \ref{fig:extract}). The ice maps of CO and OCN$^-$ found from this wavelength region are studied by Tyagi et al. (in preparation) for the IPA sources. Given the uncertainty on continuum identification of IRAS 16253, we do not analyze this object further. The band strength used for OCN$^-$ to calculate the column density is taken from \cite{Broekhuizen2004} as $1.3 \times 10^{-16}$\,cm per molecule, and the column densities are presented in Table \ref{tab:results}. The uncertainties are estimated at the ${\sim}30\%$ level which is dominated by uncertainty on the band strength.

%For example, checking the spectrum extracted from the same-size aperture on-source clearly shows detection of the OCN$^-$ feature for IRAS 16253.

\section{Ratios of icy cyanides and their comparison with gas-phase}
\label{sec:comp_gas}

\begin{figure*}
    \centering
    \includegraphics[width=0.9\textwidth]{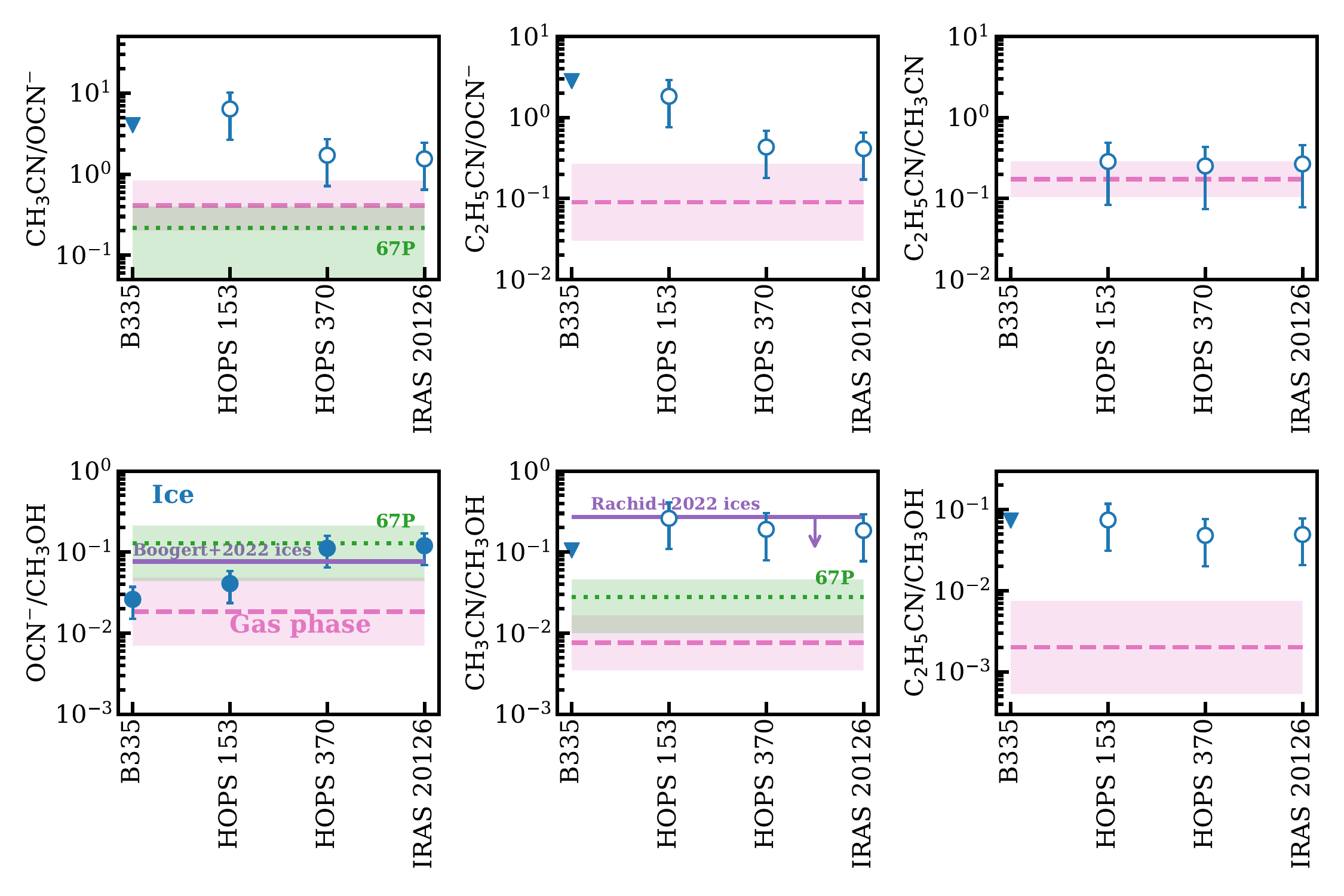}
    \caption{Ice column density ratios (blue data points) of the species studied in the this work with respect to each other and methanol taken from Slavicinska et al. (in preparation). Filled circles show firm detections, empty ones present the tentative column densities, and filled triangles show the upper limit values. The gas-phase column density ratios of many sources from the literature are shown in the pink shaded areas (the mean and standard deviation are taken from \citealt{Nazari2022ALMAGAL}). Here for the gas-phase observations we substitute OCN$^-$ by HNCO because OCN$^-$ is thought to turn into HNCO once sublimated. The purple lines show the previously measured ice column density ratios (median of values from \citealt{Boogert2022}) or upper limits (\citealt{Rachid2022}) in other sources. The abundance ratios of CH$_3$CN/CH$_3$OH, CH$_3$CN/HNCO, and HNCO/CH$_3$OH for comet 67P are also shown in green (\citealt{Rubin2019}). See Fig. \ref{fig:ratios_corr} for similar plots, but corrected for potential differences in emitting areas of the N-bearing species and methanol.} 
    \label{fig:ratios}
\end{figure*}

Figure \ref{fig:ratios} presents the column density ratios of the considered species with respect to each other and their comparison with gas-phase observations. Our ice OCN$^-$/CH$_3$OH ratios (Fig. \ref{fig:ratios} bottom left) are consistent with the median of $0.08^{+0.16}_{-0.07}$ found for 20 sources by  \cite{Boogert2022}. Moreover, our tentative column densities and upper limits for CH$_3$CN/CH$_3$OH (Fig. \ref{fig:ratios} bottom middle) agree well with the average upper limits of 0.27 found in ices for four protostellar systems by \cite{Rachid2022}. An intriguing finding is that the column density ratio of CH$_3$CN/OCN$^-$ in ices is around 1 (Fig. \ref{fig:ratios} top left). The main nitrogen ice carriers so far are thought to be NH$_3$ and NH$_4^+$ with a lesser contribution from OCN$^-$ (e.g., \citealt{Oberg2011}; \citealt{Boogert2015}). Although not sufficient to solve the missing nitrogen problem (\citealt{Pontoppidan2014}; \citealt{Altwegg2019}), our results suggest that CH$_3$CN could be an important reservoir of nitrogen in the ices as well. 

Comparing with comet 67P, our ice OCN$^-$/CH$_3$OH ratios agree within the uncertainties with HNCO/CH$_3$OH ratios in the comet (Fig. \ref{fig:ratios} bottom left). However, for CH$_3$CN/OCN$^-$ (top left) and CH$_3$CN/CH$_3$OH (bottom middle) our ice ratios are a factor of ${\gtrsim}2$ larger than the comet values of CH$_3$CN/HNCO and CH$_3$CN/CH$_3$OH. Given the uncertainties, the significance of  this difference is unclear.

Comparing with the gas-phase abundances, our ice ratios in the top row of Fig. \ref{fig:ratios} show a better agreement with gas-phase abundances than those in the bottom row. The tentative ice ratios of C$_2$H$_5$CN/CH$_3$CN (top right of Fig. \ref{fig:ratios}) for our three sources are in particularly good agreement with the gas-phase observations at a ratio of ${\sim}0.1$. This could highlight their similar origin and formation through connected chemical formation pathways. The ratios with respect to methanol (bottom row of Fig. \ref{fig:ratios}), OCN$^-$/CH$_3$OH, CH$_3$CN/CH$_3$OH, and C$_2$H$_5$CN/CH$_3$OH, show a large difference between the gas and ice (around a factor of 5 and up to a factor of ${\sim}20$).

%Since the methanol column densities used here (Slavicinska et al. in preparation) are obtained from the ${\sim}3.5$\,$\mu$m band rather than at ${\sim}9.8$\,$\mu$m where silicate could lead to saturation effects, the reason for the discrepancy cannot be the silicate band. 

These differences could be due to chemical effects, physical effects, or a combination of both. From a chemical perspective, given that the ratios with methanol seem to be most different, the chemistry of gas-phase methanol may require further study. For OCN$^-$/CH$_3$OH, we have assumed that all of OCN$^{-}$ turns into HNCO (\citealt{Oberg2009HNCO}), however, it is possible that not all the OCN$^{-}$ in ices turns into gas-phase HNCO. For example, the efficiency of conversion of OCN$^-$ to HNCO was not clear in the lab experiments of \cite{JimenezEscobar2014}. Another possibility is that OCN$^-$ could be trapped in ammonium salts as suggested to exist on comet 67P (\citealt{Poch2020}). Based on laboratory experiments those salts will then desorb at higher temperatures than methanol (${\sim}200$\,K under laboratory conditions) and during the desorption these salts can decompose into HNCO and other species (\citealt{Ligterink2018}). This difference in sublimation temperatures of OCN$^-$ and CH$_3$OH can affect the column density ratios as explained further in the following paragraph.

%Figure \ref{fig:ratios} presents the column density ratios of the considered species in the ices compared with those in the gas. Because it is thought that OCN$^-$ turns into HNCO once sublimated, here we substitute OCN$^-$ by HNCO in the gas-phase observations. 

%The ice methanol column densities for the IPA sources are taken from Slavicinska et al. (in preparation).
%Because it is thought that OCN$^-$ turns into HNCO once sublimated, here we substitute OCN$^-$ by HNCO in the gas-phase observations.

In terms of physical effects, \cite{Nazari2021} showed that if two molecules have different sublimation temperatures (different snowline locations), the gas column density ratios will not represent the true ice abundance ratios (see their Sect. 4.3). This is because normally a single source size is assumed for all compact species when calculating the column densities in unresolved gas-phase protostellar observations (e.g., \citealt{Jorgensen2018}; \citealt{Yang2021}; \citealt{vanGelder2022}). Therefore, a correction factor is needed to be multiplied by the gas-phase column density ratios which can be as large as a factor of ${\sim}10$ if the two molecules in the ratio have sufficiently different sublimation temperatures (e.g., CH$_3$OH, ${\sim}100$\,K, and NH$_2$CHO, ${\sim}300$\,K, also see Fig. 5 in \citealt{Nazari2023}). The binding energies and sublimation temperatures of CH$_3$CN, C$_2$H$_5$CN, and CH$_3$OH in the amorphous water ice matrix are similar ($T_{\rm sub}{\sim}100$\,K; \citealt{Penteado2017}; \citealt{Minissale2022}; \citealt{Ligterink2023}), so the correction factor is negligible. However, if these molecules are part of a different ice matrix the binding energies change. Thus a more direct measurement of the temperature that they trace is based on the excitation temperatures ($T_{\rm ex}$) of the gas phase observations. The $T_{\rm ex}$ for the bulk of gas-phase observations in Fig. \ref{fig:ratios} (pink shaded area) for CH$_3$CN and HNCO is around 150\,K while for methanol it is around 100\,K (Fig. 6 of \citealt{Nazari2022ALMAGAL}). Hence the correction factor to be applied to the gas-phase observations with respect to methanol is expected to be $(150/100)^{3.75}=4.6$ based on a spherical toy model (see Eq. 4 in \citealt{Nazari2021}).       

Including a correction factor of 4.6 in the gas-phase abundances of the bottom row (see Fig. \ref{fig:ratios_corr}) results in the gas-phase observations of OCN$^-$/CH$_3$OH matching the ice observations within a factor of a few. Moreover, the gas-phase observations of CH$_3$CN/CH$_3$OH and OCN$^-$/CH$_3$OH, which previously did not match the comet 67P ratios of CH$_3$CN/CH$_3$OH and HNCO/CH$_3$OH, now match the comet ratios (\citealt{Rubin2019}). Considering the uncertainties and the small sample size it is not possible to argue whether there is still a difference between the gas and ice observations of CH$_3$CN and C$_2$H$_5$CN ratios with respect to methanol. We conclude that when the physical effects are considered, the ice- and gas-phase observations are in much closer agreement with each other.

Finally, to confirm the detection and column density measurements of this work (i.e., detection of at least two ice bands for each molecule), both NIRSpec and MIRI data of these sources should be analyzed in the future. Figure \ref{fig:lab_spectra_CN} presents the laboratory spectrum of pure CH$_3$CN and C$_2$H$_5$CN in the $2.5-12.5$\,$\mu$m regime. We note the CH$_3$CN feature at 3.33\,$\mu$m, which has a smaller band strength than the feature analyzed here, falls in the bulk stretch of the H$_2$O ice band and thus is challenging to detect. After the strongest feature at 4.44\,$\mu$m, the next best features are in the wavelength range of MIRI (${\sim}6.8-7.4$\,$\mu$m). Those are particularly promising because of their larger FWHM and thus their larger or similar band strengths (\citealt{Rachid2022}). Although those features are embedded in strong ice features of other simple and complex species (\citealt{Yang2022}; \citealt{Rocha2023}), they could provide an additional constraint on the column densities of complex cyanides.

\begin{figure}
    \centering
    \resizebox{\hsize}{!}{\includegraphics{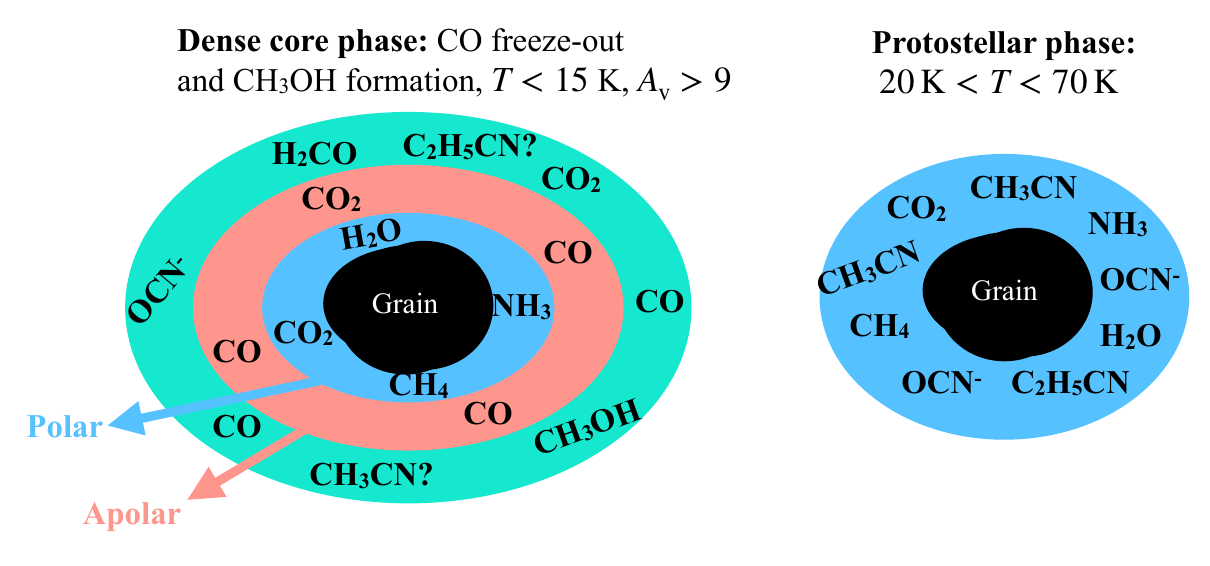}}
    \caption{Sketch of typical dust grain chemical inventory. We suggest that OCN$^{-}$, CH$_3$CN and C$_2$H$_5$CN likely start to form in the cold dense core phase but are enhanced thermally during the warm-up stage where more volatile species have sublimated.}
    \label{fig:story}
\end{figure}

\section{Chemical interpretation} 
\label{sec:chem}

\subsection{Origin of complex cyanides}
OCN$^-$ is thought to start forming in the dense core phase after the CO freeze-out (see the review by \citealt{Boogert2015}). Because the CH$_3$CN:CO mixture was needed to fit the data, we speculate that CH$_3$CN starts forming in this same phase along with OCN$^-$ (Fig. \ref{fig:story}). In fact, this agrees with the constant gas-phase column density ratios of CH$_3$CN/CH$_3$OH with some low-level scatter found for a large sample of protostars by \cite{Nazari2022ALMAGAL}, which points to their formation likely in the cold prestellar phase. This interpretation also agrees with our finding that gas and ice ratios are similar to each other after correction for source structure. The other important ice mixture explaining the data is CH$_3$CN:H$_2$O:CO$_2$, which is expected if the ices are thermally processed (see Sect. \ref{sec:thermal}). We emphasize that the spectral fits presented here (Fig. \ref{fig:fits}) are not unique, however, mixtures of CH$_3$CN with CO or NH$_3$, CO$_2$ and H$_2$O are always needed. Because there are no laboratory spectra of C$_2$H$_5$CN ice mixtures, we cannot constrain its ice environment. The simplest assumption is that it also starts forming with OCN$^-$ and CH$_3$CN in the dense core phase.

\subsection{Thermal processing}
\label{sec:thermal}

\begin{figure}
  \resizebox{\hsize}{!}{\includegraphics{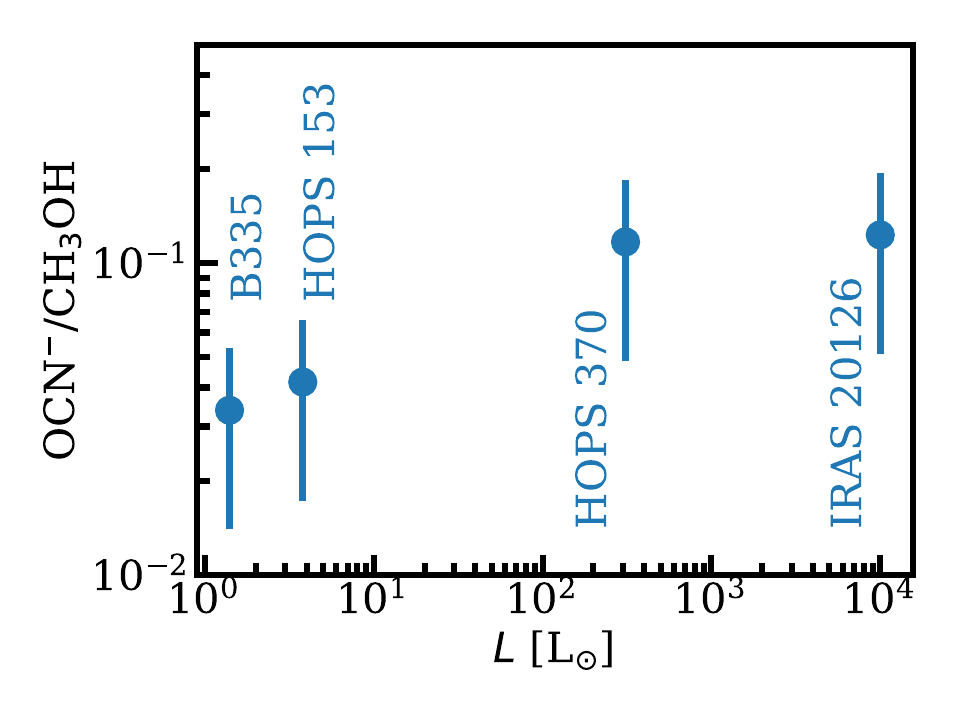}}
  \caption{Column density ratios of OCN$^-$/CH$_3$OH as a function of luminosity for the IPA sources.}
  \label{fig:OCN_L}
\end{figure}

Clues on the formation pathways are obtained by considering the relation between column density ratios and the bolometric luminosity. The OCN$^-$/CH$_3$OH ratios increase slightly with luminosity for our sample (Fig. \ref{fig:OCN_L}). Additional sources are needed to confirm this relation. The OCN$^-$ ice feature is enhanced at higher temperatures in laboratory experiments (e.g., \citealt{Novozamsky2001}; \citealt{Broekhuizen2004}), and a potential correlation between OCN$^-$ and luminosity is justified if the sources with higher luminosities also have higher ice temperatures with more volatile species sublimated. Brunken et al. (submitted) use spectra extracted on-source and find that IPA sources with higher bolometric luminosities also show a more prominent $^{13}$CO$_2$ double-peaked ice feature, which is known to indicate ice thermal processing. The locations where we extracted the spectra still show a prominent $^{13}$CO$_2$ double-peaked feature for IRAS 20126 and HOPS 370 (Fig. \ref{fig:optical_depth}), while the double-peaked feature is less prominent (or non-existent) for the other sources. Therefore, ices of IRAS 20126 and HOPS 370 (the two most luminous sources in the sample) are warmer than the other sources and thus, OCN$^-$ formation is likely enhanced (Fig. \ref{fig:story}).

It is interesting to note that the three sources with a tentative detection of CH$_3$CN and C$_2$H$_5$CN are those with the highest luminosities and potentially warmest ices. Moreover, the laboratory CH$_3$CN:CO ice mixture that fitted the data best were those at high temperatures. Hence formation of CH$_3$CN and C$_2$H$_5$CN might be enhanced in warmer ices. However, the non-detection in the other two sources could be simply a sensitivity problem. Therefore, a larger sample size in which these ices are detected with a range in luminosities and masses is needed to further study thermal processing of CH$_3$CN and C$_2$H$_5$CN in interstellar ices.

Nevertheless, this potential enhancement of CH$_3$CN and C$_2$H$_5$CN formation at higher temperatures is expected from their ice formation pathways. In the ice, CH$_3$CN is thought to form through reaction of two radicals CH$_3$ and CN (\citealt{Garrod2008}; \citealt{Garrod2022}). Because ice radicals can have higher mobility at higher temperatures, CH$_3$CN production can increase. For C$_2$H$_5$CN, laboratory experiments using UV photolysis (\citealt{Bulak2021}) found that higher temperatures enhance the formation of C$_2$H$_5$CN from CH$_3$CN-based ices. Enhanced formation of CH$_3$CN and C$_2$H$_5$CN in warmer ices can also help explain the low-level scatter in the column density ratios of gas-phase observations (pink area in Fig. \ref{fig:ratios}). 

%To conclude, we are likely witnessing the transition from desorption of CO to the mixing of OCN$^-$, CH$_3$CN and C$_2$H$_5$CN with H$_2$O and CO$_2$ ices during the warm-up stage. 

\section{Conclusions}
\label{sec:conclusion}

In this paper we report the first tentative detection of CH$_3$CN, C$_2$H$_5$CN, and N$_2$O in ices around three embedded protostars (HOPS 153, HOPS 370, and IRAS 20126+4104). Our CH$_3$CN/OCN$^-$ ice ratios are on the order of 1, suggesting that CH$_3$CN could be a significant nitrogen reservoir in ices. Our ice column density ratios among the nitrogen-bearing species agree better between the gas and ice observations in comparison to those with respect to methanol which are around a factor of 5 larger in ices than the gas-phase. We attribute this to the difference in the snowline locations of CH$_3$OH and the nitrogen-bearing species studied here. 

The main ice mixtures that fit the data are those where CH$_3$CN is mixed with H$_2$O, CO$_2$ and CO. We conclude that CH$_3$CN and C$_2$H$_5$CN likely start forming in the dense core phase after the CO freeze-out and could be further enhanced during the warm-up stage around the protostar. 

\textit{JWST}-NIRSpec data presented here help in understanding the chemistry and formation mechanism of complex cyanides. However a much larger sample of protostars with both NIRSpec and MIRI data are needed to firmly identify and statistically measure the significance of the relations proposed here.   

%This is especially important because planets could start forming in the early protostellar phase from icy dust grains which contain these molecules (\citealt{Tychoniec2018}; \citealt{Tobin2020}; \citealt{Cridland2022}) or these COMs can be delivered to planets by comets at later stages (see \citealt{Altwegg2019} for review). 

%This is especially relevant, given that planets could start forming in the early protostellar phase (\citealt{Tychoniec2018}; \citealt{Tobin2020}; \citealt{Cridland2022}) or N-bearing COMs can be delivered to planets by comets at later stages (see \citealt{Altwegg2019} for review).

\begin{acknowledgements}
    We thank the referee for their constructive comments. P.N. greatly appreciates the helpful discussions with Danial Langeroodi. Astrochemistry in Leiden is supported by the Netherlands Research School for Astronomy (NOVA), by funding from the European Research Council (ERC) under the European Union’s Horizon 2020 research and innovation programme (grant agreement No. 101019751 MOLDISK), and by the Dutch Research Council (NWO) grant 618.000.001. Support by the Danish National Research Foundation through the Center of Excellence “InterCat” (Grant agreement no.: DNRF150) is also acknowledged. Support for AER, STM, RG, DW, and SF in program \#1802 was provided by NASA through a grant from the Space Telescope Science Institute, which is operated by the Association of Universities for Research in Astronomy, Inc., under NASA contract NAS 5-03127. G.A. and M.O. acknowledge financial support from grants PID2020-114461GB-I00 and CEX2021-001131-S, funded by MCIN/AEI/10.13039/501100011033. Y.-L.Y. acknowledges support from Grant-in-Aid from the Ministry of Education, Culture, Sports, Science, and Technology of Japan (20H05845, 20H05844, 22K20389), and a pioneering project in RIKEN (Evolution of Matter in the Universe). This work is based on the observations with the NASA/ESA/CSA James Webb Space Telescope. The data were obtained from the Mikulski Archive for Space Telescopes at the Space Telescope Science Institute, which is operated by the Association of Universities for Research in Astronomy, Inc., under NASA contract NAS 5-03127 for JWST. These observations are associated with program \#1802. All the JWST data used in this paper can be found in MAST: \href{https://archive.stsci.edu/doi/resolve/resolve.html?doi=10.17909/3kky-t040}{10.17909/3kky-t040}. 
\end{acknowledgements}


\bibliographystyle{aa}
\bibliography{ice_CH3CN}

\begin{appendix}

\section{Additional tables and plots}

\begin{figure*}
    \centering
    \includegraphics[width=0.9\textwidth]{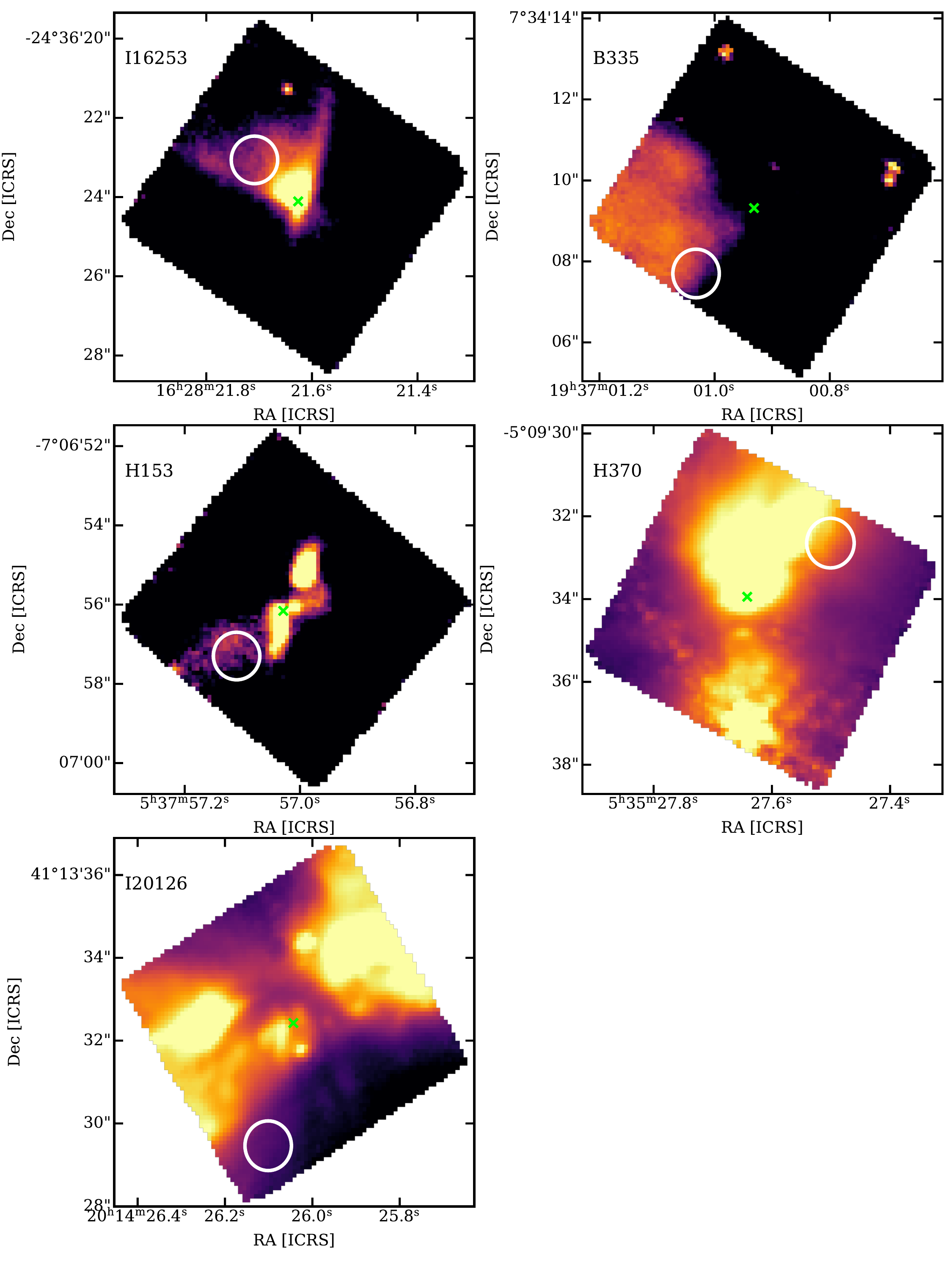}
    \caption{\textit{JWST}-NIRSpec continuum images at 3.92\,$\mu$m for the IPA targets. White circles show the apertures from which the spectra are extracted in this work to avoid the CO rotational-vibrational emission lines as much as possible. The central position of each protostar according to the IPA-MIRI channel 3 medium band is shown by a green cross.} 
    \label{fig:extract}
\end{figure*}

\begin{figure*}
    \centering
    \includegraphics[width=0.9\textwidth]{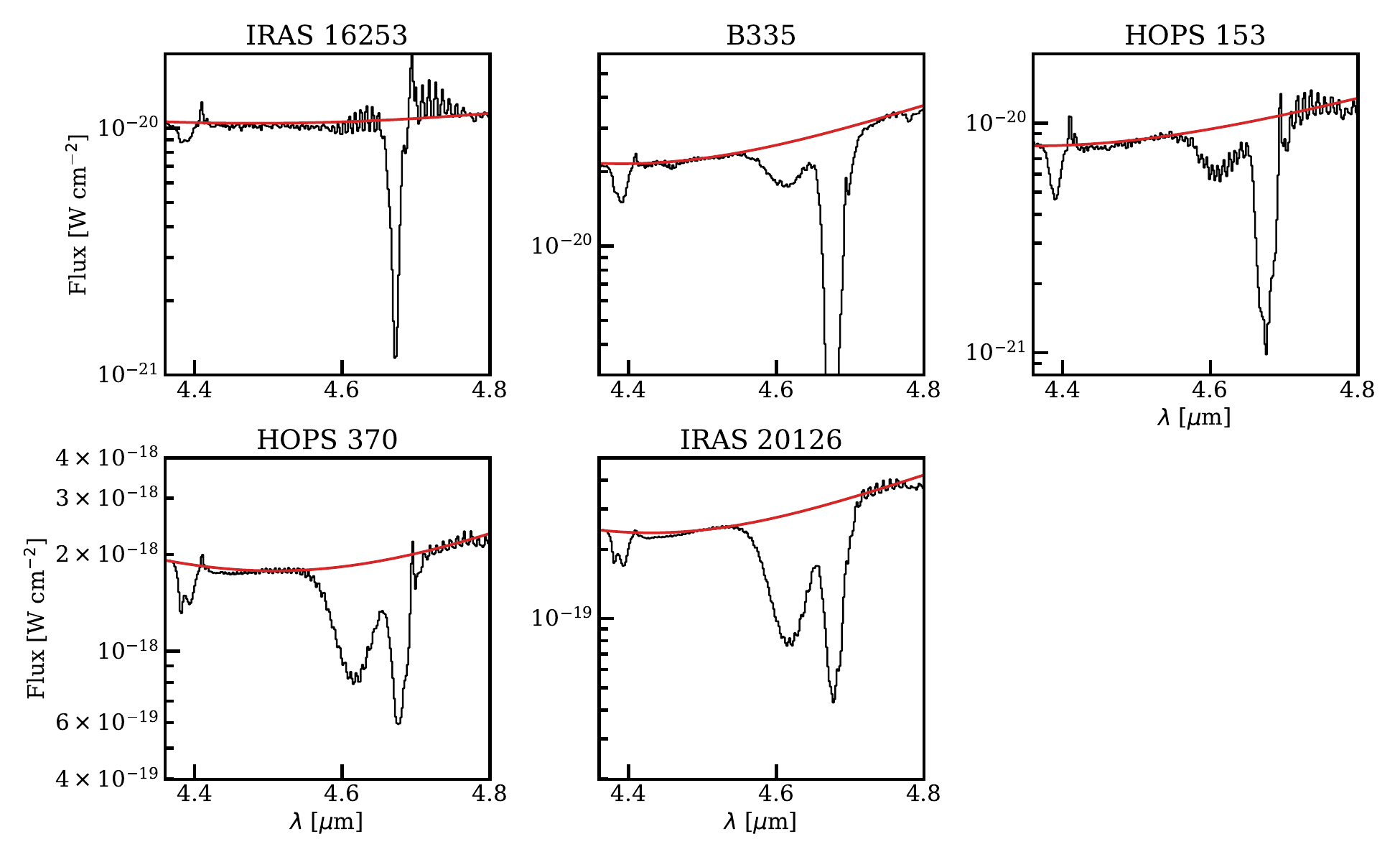}
    \caption{Spectra (black) and the fitted continuum (red) for all the five sources in the IPA program. The continuum is fitted locally using a polynomial function.} 
    \label{fig:cont}
\end{figure*}

\begin{figure*}
    \centering
    \includegraphics[width=0.9\textwidth]{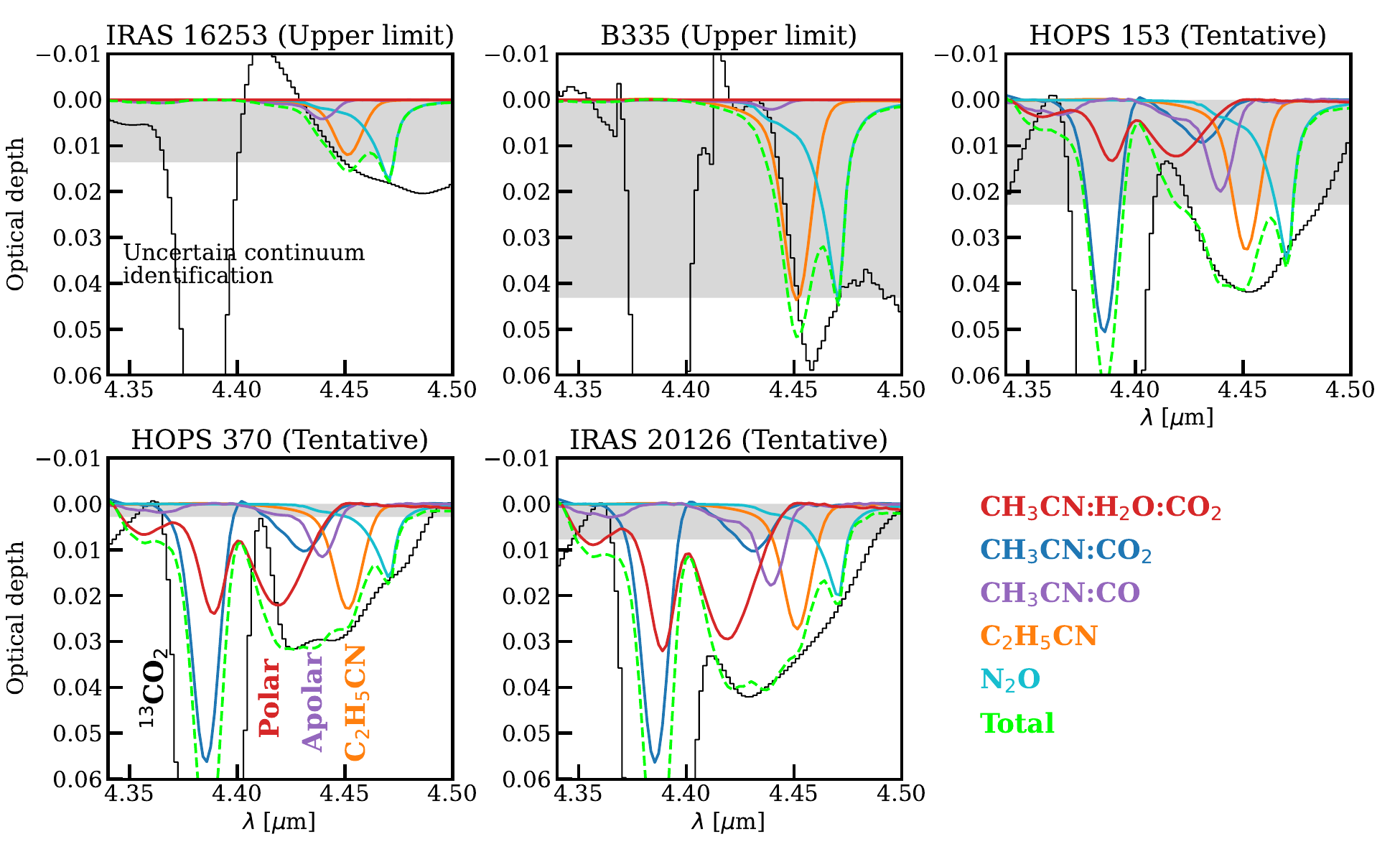}
    \caption{Same as Fig. \ref{fig:fits} but for the emission-line-subtracted spectra taken from \cite{Rubinstein2023} where the methods of emission line fitting and subtraction are explained.} 
    \label{fig:fit_baseline}
\end{figure*}

\begin{figure*}
    \centering
    \includegraphics[width=0.9\textwidth]{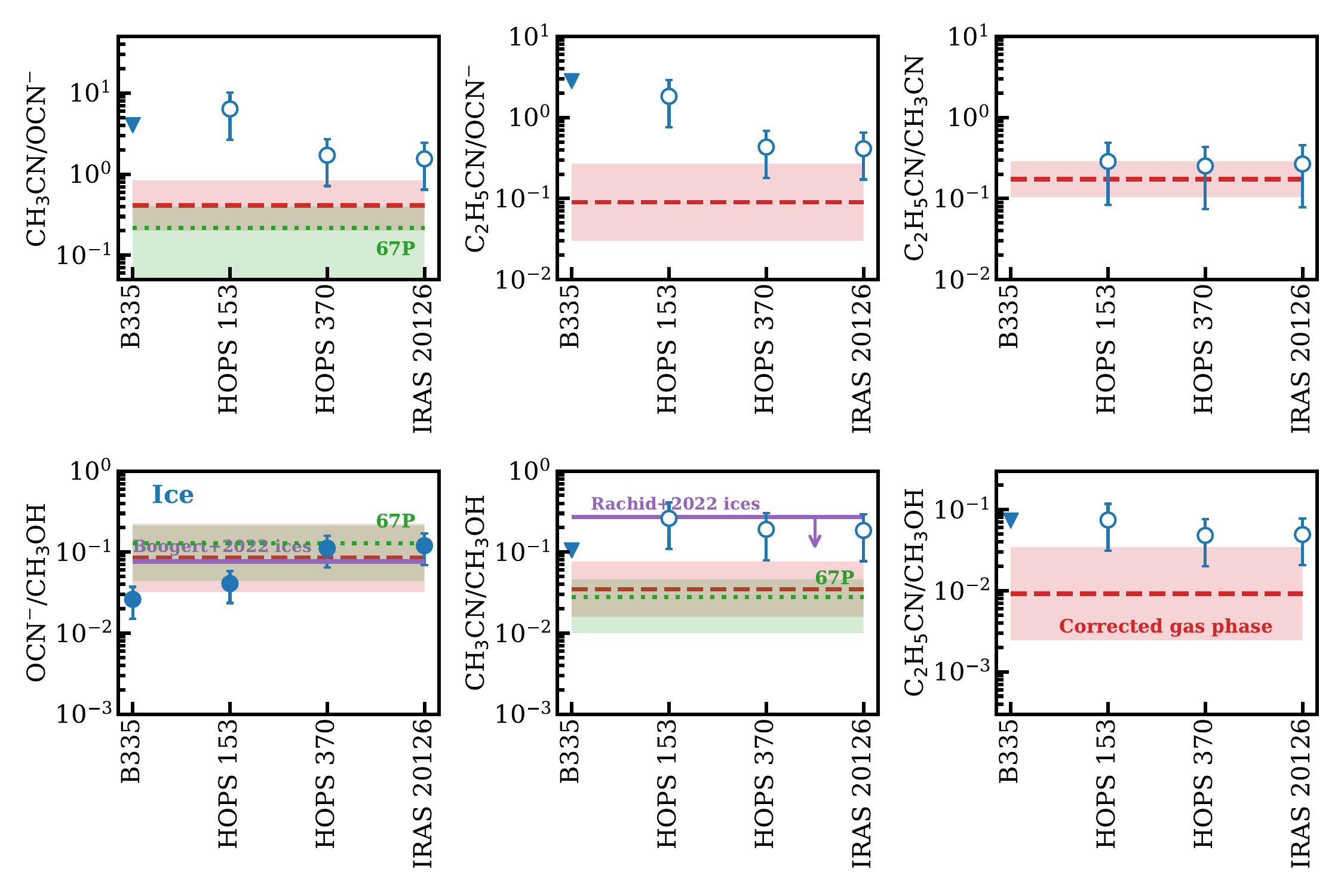}
    \caption{Same as Fig. \ref{fig:ratios} but now the gas-phase column density ratios are corrected by a factor of 4.6 for potential differences between the snowline locations of nitrogen-bearing molecules with respect to methanol.} 
    \label{fig:ratios_corr}
\end{figure*}

\begin{figure*}
    \centering
    \includegraphics[width=0.9\textwidth]{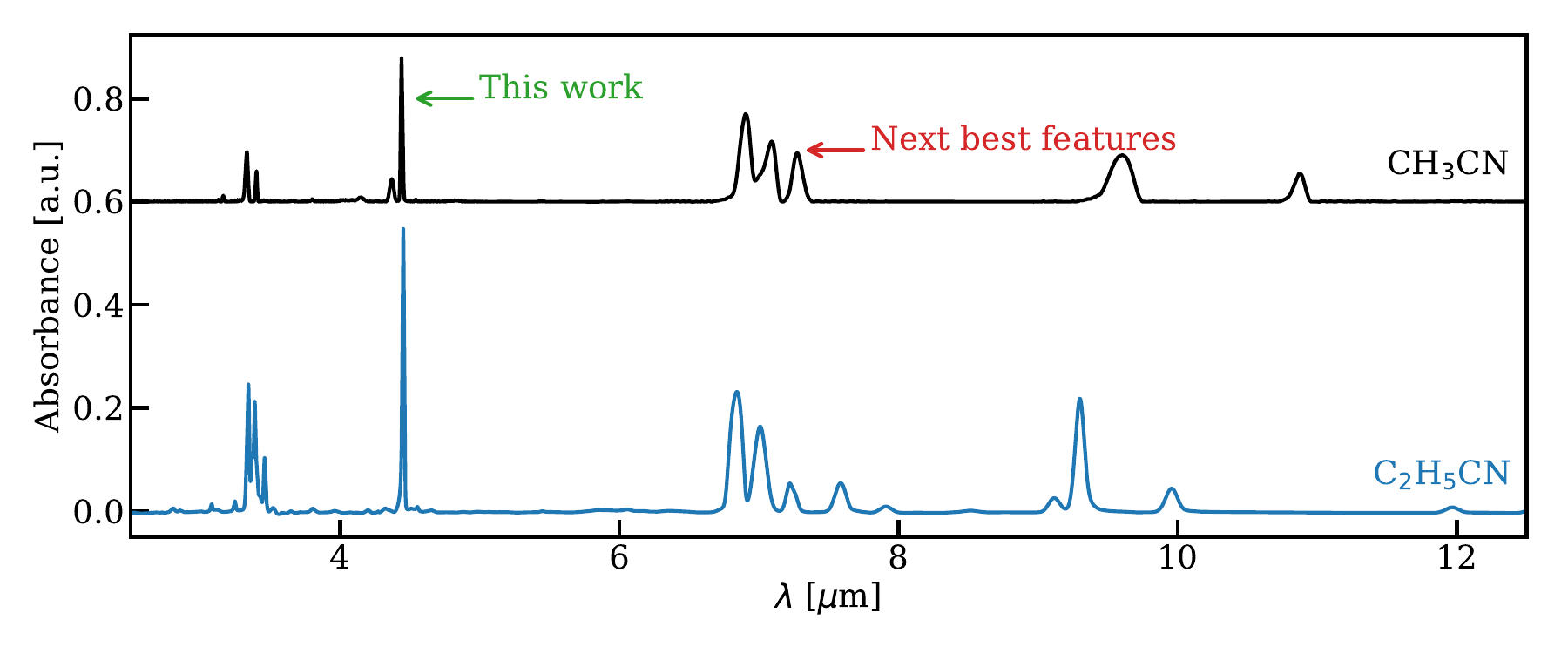}
    \caption{Laboratory spectra of pure CH$_3$CN and C$_2$H$_5$CN at 15\,K and 50\,K, respectively, to highlight which feature can be used in the future for firm detection of these two species. The spectrum of CH$_3$CN is shifted vertically for better readability.} 
    \label{fig:lab_spectra_CN}
\end{figure*}

Table \ref{tab:obs} presents the centers of the apertures from which the spectra are extracted, in addition to the rms on the continuum optical depth and the source characteristics. Table \ref{tab:band_strengths} presents the band strengths used for the column density measurements. Figure \ref{fig:extract} illustrates the positions from which the spectra are extracted in this work. Figure \ref{fig:cont} presents the local continuum fits to the extracted spectra. Figure \ref{fig:fit_baseline} presents the fit to the CN-stretching region for the emission-line-subtracted spectra (see \citealt{Rubinstein2023}). Figure \ref{fig:ratios_corr} is the same as Fig. \ref{fig:ratios} but for corrected gas-phase column density ratios. Figure \ref{fig:lab_spectra_CN} presents the laboratory spectra of pure CH$_3$CN and C$_2$H$_5$CN to highlight the strongest features. 

%Figure \ref{fig:OCN_L} presents the ice column density ratios of OCN$^{-}$/CH$_3$OH as a function of luminosity. 

\begin{table*}
%\Huge
\centering
\renewcommand{\arraystretch}{1.3}
    \caption{Positions for spectral extraction and source properties}
    \label{tab:obs}
    \begin{tabular}{lllllll}
          \toprule
          \toprule    
        Source & R.A. & Decl. &Optical depth  & $L_{\rm bol}$& $d$ & References\\
        &[hh:mm:ss] & [dd:mm:ss] & rms& [L$_{\odot}$] & [pc] &\\
        \midrule     

IRAS 16253&16:28:21.71&-24:36:23.06&0.0045 & 0.16 &  140 & \cite{Ortiz2018IPA}; \cite{Aso2023}\\
B335&19:37:01.03&07:34:07.70&0.014 & 1.4&  165 & \cite{Watson_2020}; \cite{Evans2023}\\
HOPS 153&05:37:57.11&-07:06:57.30&0.0076 & 3.8 &  390 & \cite{Tobin2020}  \\
HOPS 370&05:35:27.50&-05:09:32.65&0.0009 & 315.7&  390 & \cite{Tobin2020} \\
IRAS 20126&20:14:26.10&41:13:29.46&0.0026 & 10$^4$& 1550 & \cite{Johnston2011}; \cite{Reid2019}\\

\bottomrule
        \end{tabular}
        \tablefoot{Second and third column give the right ascension and declination of the center of the aperture used to extract the spectrum (see Fig. \ref{fig:extract}). The references of bolometric luminosities and distances of the central sources are given in the right-most column.}
\end{table*}

\begin{table}
\tiny
\renewcommand{\arraystretch}{1}
    \caption{Band strengths}
    \label{tab:band_strengths}
    \resizebox{0.7\columnwidth}{!}{\begin{tabular}{@{\extracolsep{0.2mm}}*{2}{l}}
          \toprule
          \toprule    
        Ice species & $A$\,(cm\,molecule$^{-1}$) \\
        \midrule     

CH$_3$CN:CO$_2$ & $1.5 \times 10^{-18}$ \\
CH$_3$CN:CO & $1.5 \times 10^{-18}$ \\
CH$_3$CN:H$_2$O:CO$_2$ & $3.3 \times 10^{-18}$ \\
C$_2$H$_5$CN & $2.9 \times 10^{-18}$\\
OCN$^-$ & $1.3 \times 10^{-16}$ \\
\bottomrule
        \end{tabular}}
        \tablefoot{The band strengths are taken from \cite{Rachid2022} and \cite{Broekhuizen2004} for CH$_3$CN mixtures and OCN$^-$, respectively, while they are calculated for C$_2$H$_5$CN in this work.}
\end{table}

% \begin{figure}
%   \resizebox{\hsize}{!}{\includegraphics{diff.pdf}}
%   \caption{The spread factor of gas-phase observations as a function of the difference between ice observations and the gas. The spread factor is defined as 10$^{\rm dex}$ and presents the level of scatter in gas-phase observations (taken from \citealt{Nazari2022ALMAGAL}). The x-axis presents the difference between ($log_{10}$ of) the mean of the ratios for the tentatively detected ice values (or the median of ice ratios from \citealt{Boogert2022} for OCN$^-$/CH$_3$OH) with ($log_{10}$ of) the mean of ratios in gas-phase observations.}
%   \label{fig:diff}
% \end{figure}

\end{appendix}

\end{document}